%% file: mc_ipu.tex
\documentclass[pdflatex,sn-nature]{sn-jnl}
\usepackage{graphicx}%
\usepackage{multirow}%
\usepackage{amsmath,amssymb,amsfonts}%
\usepackage{amsthm}%
\usepackage{mathrsfs}%
\usepackage[title]{appendix}%
\usepackage{xcolor}%
\usepackage{textcomp}%
\usepackage{manyfoot}%
\usepackage{booktabs}%
\usepackage{siunitx}
\usepackage{nicematrix}
\usepackage{array}
\usepackage{natbib}

\begin{document}

\title{Fast and Flexible Spiking Network Simulations on Massively Distributed On-Chip Memory Accelerators}

\author*[1,2]{\fnm{Catherine M.} \sur{Sch\"ofmann}}\email{c.schoefmann@fz-juelich.de}
\author[1,2]{\fnm{Jan} \sur{Finkbeiner}}
\author[1,2]{\fnm{Jan} \sur{Vogelsang}}
\author[1]{\fnm{Susanne} \sur{Kunkel}}

\affil[1]{\orgdiv{Neuromorphic Software Ecosystems (PGI-15)}, \orgname{Jülich Research Centre}, \orgaddress{\city{Jülich}, \country{Germany}}}
\affil[2]{\orgname{RWTH Aachen University}, \orgaddress{\city{Aachen}, \country{Germany}}}

\abstract{
Studying brain function and dysfunction in spiking networks with realistic synaptic densities pushes simulation technology beyond its current limits. It requires simulations substantially faster than biological real-time to cover long model times and enable efficient iterative exploration, particularly when connectivity is plastic. The modelling community aspires to a 100× speedup. Yet on conventional many-core systems this remains elusive, bounded by inter-node communication and memory access. So far, significant speedup has only been shown on a custom-built FPGA system. We demonstrate the potential of commercial systems with large distributed on-chip memory to accelerate brain research. To exploit the local SRAM, we introduce a group-based mapping and communication strategy tackling spike-routing challenges on massively distributed systems. On Graphcore IPUs, the cortical microcircuit runs at a record $22.7\times$ faster than biological real-time with negligible build times and a reusable compiled graph. By design, the algorithm makes no model- or hardware-specific assumptions, providing a general strategy for advancing neuronal simulation technology.}

\maketitle

\section{Introduction}
\label{sec:intro}

\begin{figure}
\centering
\includegraphics[width=0.92\textwidth]{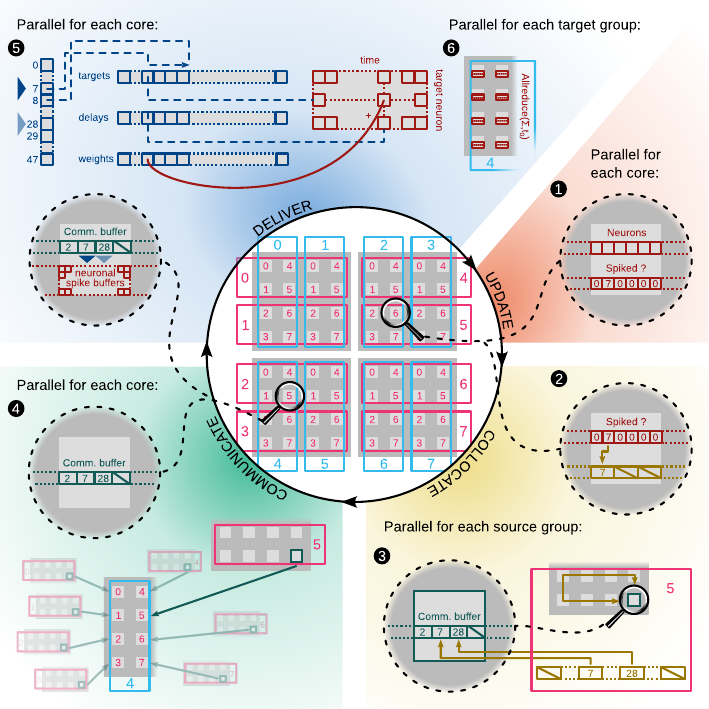}
\caption{\textbf{Group-based algorithm.} Centre: the cores (small grey squares) of all compute nodes (large grey squares) form a grid of source groups (numbered pink rows) and target groups (numbered cyan columns). The pink number on each core denotes the source group whose synapses it stores. Surrounding panels show
one simulation cycle. \emph{Update} (red): cores advance their neurons by one time step and record which spike. \emph{Collocate} (yellow): the source group
local IDs of spiking neurons are compacted into one buffer per source group, with the spike count in the first slot. \emph{Communicate} (green): each source group's buffer is sent to its matching core in every target group. \emph{Deliver} (blue): each core looks up the synapses of every received ID, adds their weights to its copy of the target group's spike buffer at the slot given by the delay, and the copies are summed within the target group.}
\label{fig:overview}
\end{figure}

Comprehensive simulation studies of cortical networks depend on compute systems capable of running models significantly faster than biological real-time. Any improvement in simulation efficiency directly benefits neuroscientists in modelling tasks, allowing for more extensive parameter scans and larger scale experiments without the need for massive clusters. For example, capturing learning processes such as long-term plasticity requires tracking network activity over biological hours or days~\cite{learningTakesLong}, demanding simulation speeds that remain difficult to achieve at scale on current hardware.
For a large variety of neural simulations, CPUs remain the most used target platform~\cite{diesmann2001nest,hines1997neuron,learningTakesLong}. Increasingly efficient algorithm design~\cite{jordan2018extremely} has led to a consistent increase of simulators' speedup relative to biological simulated time $T_\mathrm{bio}$ (real-time factor, $\mathrm{RTF}=T_\mathrm{wall}/T_\mathrm{bio}$) over the years. Across CPU, GPU and SpiNNaker implementations of the same benchmark, however, these gains largely follow hardware improvements. The achieved RTF approximately scales with the square of the process node and linearly with the energy per synaptic event, the only exception being FPGA-based systems~\cite{Senk_2026}. One major bottleneck limiting further speedups lies in the very nature of these simulations: low compute requirements with frequent and often irregular memory access, along with unstructured communication needs between neuron-hosting processes. Due to the separation of memory and compute in the von-Neumann architecture, runtimes are dominated by cores waiting for data movements to complete. This is especially true for simulations of point-neuron models, since the computational load for simple update equations (e.g. Leaky Integrate-and-Fire, LIF) is minimal. GPU-based simulators~\cite{knight2021pygenn,neuronGPU} offer a more compact compute-memory layout than CPUs, enabling slightly more graceful scaling, but ultimately suffer from the same problem. Moreover, especially larger neural models tend to exhibit extremely sparse connectivity. A single neuron is unlikely to have a large number of post-synaptic connections on any given process~\cite{jordan2018extremely}, leading to spike delivery suffering the most from disjoint and uncoalesced memory access.

The key to breaking this scaling barrier, therefore, lies in efficient routing. A recent breakthrough was made by the FPGA-based neuroAIx simulator~\cite{kauth2023neuroaix}, which uses a dedicated router unit and algorithm to simulate the cortical microcircuit benchmark at $20\times$ biological real-time. This comes with flexibility trade-offs typical of FPGA-based implementations, where adopting a different neuron model or network often requires re-synthesising the design. We aim to further improve on this level of performance without trading off flexibility.

State-of-the-art technology for distributed simulations of spiking networks on conventional hardware typically advances in discrete time steps and periodically exchanges spikes between compute nodes to preserve causality~\cite{Morrison05a}. The interval over which each node may progress independently is set by the minimum synaptic transmission delay; one such interval constitutes a simulation cycle. In line with NEST terminology (e.g.,~\cite{Lober_2026}), we distinguish four phases. In the \emph{update} phase, every node advances the dynamical state of its local neurons over the minimum-delay interval, optionally in smaller steps, and records which of them cross the firing threshold. In the \emph{collocate} phase, the resulting spikes are compacted into communication buffers. In the \emph{communicate} phase, these buffers are exchanged between nodes through a collective operation. In the \emph{deliver} phase, the synaptic weights are added to the corresponding target neuron's spike buffer, contributing to the state update at the point in simulated time set by the remaining synaptic transmission delay. We evaluate our approach using the full-scale Potjans-Diesmann cortical microcircuit model (PD14), a standard benchmark comprising approximately \num{77000} neurons and \num{3e8} synapses (Sec.~\ref{sec:PD14})~\cite{potjans2014,Senk_2026}. 

For our solution, we make use of distributed memory architectures, systems that allocate fast volatile memory (i.e., SRAM) to every computing core instead of organising it in cache hierarchies or sharing between parallel units. This reduces memory access latency of each processor to its local memory to a small constant duration, regardless of the access pattern, offering orders-of-magnitude lower latency and higher bandwidth compared to external memory (DRAM). 
Neuromorphic systems often provide dedicated hardware support for spike communication through event-driven or mixed-signal designs, but constrain the model to fit it. SpiNNaker was designed for \SI{1}{\milli\second} time steps and fixed-point arithmetic, and required a $20\times$ slow-down to simulate PD14 at \SI{0.1}{\milli\second} resolution~\cite{spinnakerpd14}, with real time reached only through model-specific
redistribution of synaptic processing~\cite{spinnakerRealtime}. Loihi~\cite{loihiWhitepaper} and TrueNorth~\cite{akopyan2015truenorth} restrict weight resolution, delays and neuron dynamics, while the analog BrainScaleS-2 system~\cite{brainscaleS} operates with limited neuron fan-in, provides 256 synapses per neuron circuit and \num{131072} synapses per chip. PD14, with \num{3880} synapses per neuron and $\num{3e8}$ in total, would therefore require over \num{2000} chips. In parallel, machine-learning accelerators such as Cerebras~\cite{cerebrasWhitepaper}, NorthPole~\cite{northpole} and the IPU~\cite{ipuWhitepaper} demonstrate the value of colocating memory and compute at scale.  We choose the IPU as reference hardware for our implementation, as it offers a large amount of local memory per core and provides a flexible low-level software stack with its custom \texttt{poplar} SDK.

In order to make use of low access latency for local memory, we implement a parallel spike delivery algorithm that maximises local computation by structuring spike exchange between groups of cores.
Our algorithm defines communication pathways for a configurable group size at build time, thus reducing the required filtering as every post-synaptic target builds a connection lookup table for its pre-synaptic source groups ahead of simulation.
Each core is part of both a source and a target group. The source group defines which buffer its spikes will be aggregated into, while a core's position in a target group defines which source-target group pair it will hold the synapse data for. By tweaking the ratio of groups, we can change the communication granularity and which stages of the simulation benefit most from parallelisation. The optimal configuration will depend on the hardware and implementation of inter-core communication.

Combined with the fast random access times of distributed memory, this architecture underpins our state-of-the-art performance. At the same time, the IPU retains the flexibility of a higher-level software-programmable accelerator. Changes to the model or implementation only require recompiling C++ codelets, which takes seconds as opposed to hardware re-synthesis. We achieve simulation times $22.7\times$ faster than biological real-time, meaning a $1.11\times$ speedup and approximately $3.6\times$ better energy efficiency compared to the previous state-of-the-art neuroAIx~\cite{kauth2023neuroaix}. Part of this gap reflects that our implementation requires no external memory, but the values also differ in measurement level (IPU device power vs. FPGA board power, Sec.~\ref{sec:results_energy}). Furthermore, we demonstrate that, next to our 32-bit default, the architecture can utilise 16-bit precision synaptic weights to minimise memory consumption while still delivering accurate simulation output. Verification is done through comparison of the same model run on CPU-based simulator NEST~\cite{schofmann_2026_18268837}, see Appendix~\ref{app:verification}.

\section{Results}
\label{sec:results}

\subsection{Group-based Algorithm}
\label{sec:algooutline}
The time-driven simulation advances by repeating the same four phases every cycle (Fig.~\ref{fig:overview}): evolving the neuron states (\emph{update}), collecting emitted spikes into communication buffers (\emph{collocate}), exchanging these buffers between cores (\emph{communicate}), and delivering the spikes to their target neurons (\emph{deliver}). The work within the update, collocate, and deliver phases is parallelised, but cores must synchronise across node boundaries in the communicate phase.

The key idea of our approach is a group-based mapping of synapses to cores. Functional groups partition the cores such that each core is assigned to one source group and one target group. The $C$ cores form a $G_\mathrm{src}\times G_\mathrm{tgt}$ grid, where each source group spans $G_\mathrm{tgt}$ cores (one per target group) and vice versa each target group spans $G_\mathrm{src}$ cores (one per source group). All neurons of the simulated network are distributed round-robin across the source groups (interleaved per neuron population) and, within each source group, across its cores (see Appendix~\ref{app:loadbalance}). The groups determine the mapping of synapses to cores: a target group hosts all incoming synapses of its neurons, while each core of a target group hosts only the synapses originating from a specific source group. This mapping enables a group-based communication scheme and shapes all four phases of the simulation cycle.
In the \emph{update} phase, neuron dynamics evolve, integrating spike contributions and external input as well as detecting threshold crossings of the neuronal membrane potential. Each core is responsible for the update of its local neurons $N_C\,=\,N/C$, where $N$ is the total number of neurons in the network. The benchmark network model PD14 (Sec.~\ref{sec:PD14}) consists of simple current-based leaky integrate-and-fire neurons, allowing for exact integration of the model dynamics~\cite{rotter1999exact}. A per-neuron buffer records the emitted spikes of the current time step as the local neuron IDs of the respective source group.

In the \emph{collocate} phase, spikes are hierarchically gathered into communication buffers. A source group manages the joint communication of the spikes emitted by its corresponding $N_\mathrm{src}\,=\,N_C\,G_\mathrm{tgt}$ neurons. To this end, every neuron $j$ is assigned a local index $n_j^\mathrm{src}\in\{0,1,2,\dots,N_\mathrm{src}-1\}$. Each source group has one or multiple root cores, which gather all spikes emitted by the source group in a given time step. This is done over several intermediate aggregation steps, resulting in a final payload containing the number of spikes followed by a contiguous list of local indices.

In the \emph{communicate} phase, the packed spike buffers are exchanged between source and target groups according to the group-based mapping. Each target group receives the buffers of all source groups, while each core within the target group receives only the buffer of its corresponding source group (Fig.~\ref{fig:overview}).

In the \emph{deliver} phase, each core processes the spikes in its received buffer and routes them to their target neurons. Target group cores locally hold all synapse data for the matching source group and each core has a copy of the entire target group's ring buffer $(C,d_{\max},N_\mathrm{tgt})$, a data structure meant to hold spike weights until they can be processed by the target neuron with a simulated delay up to a maximum number of steps, $d_{\max}$. The copy is necessary to enable parallel routing of spikes. Synapse data comprises weights, delays, and the target group local index $n_j^\mathrm{tgt}\in\{0,1,2,\dots,N_\mathrm{tgt}-1\}$, which identifies the target neuron within the target group. Synapse data (weights, delays, targets) are stored per core in three contiguous arrays sorted by source. This information is accessed using an index range (begin, end) into the synapse data, retrieved from indexing into a lookup table with the incoming spikes' source group local index $n_j^\mathrm{src}$.
We then iterate over the index range and add each synapse weight in its corresponding ring buffer slot, identified by the local index $n_j^\mathrm{tgt}$ and an offset from the current time step index given by the synapse delay. Finally, each target group performs a reduce over its $G_\mathrm{src}$ ring buffers. This ensures that every target group neuron $n_j^\mathrm{tgt}$ has a complete view of its received spikes.

\subsection{Memory Usage and Workload}
\label{sec:memory_usage_workload}
Because the group mapping is independent of the instantiated connectivity, its expected per-core memory footprint can be predicted before graph construction as shown in Sec.~\ref{sec:methods_prediction_mem_usage} and Fig.~\ref{fig:performance}d). Increasing the number of source groups reduces the lookup-table footprint as each core represents fewer potential source neurons, but enlarges the target groups and therefore the neuronal spike buffers. The synapse-data contribution remains approximately constant. Measured memory usage closely follows this prediction and is minimized at an intermediate group ratio of $G_\mathrm{src}=16$. Small deviations across connectivity seeds arise because, although the connection count between each population pair is fixed, the individual pre- and post-synaptic neuron pairs are sampled randomly, producing seed-dependent differences in the synapse load assigned to each core (see Sec.~\ref{sec:PD14}). On $M=4$ IPUs, configurations with $G_\mathrm{src}=4$ and $G_\mathrm{src}\geq 64$ exceed the memory budget, as minor overhead from runtime structures (see Sec.~\ref{sec:methods_prediction_mem_usage}) pushes slightly over the predicted value in Fig.~\ref{fig:performance}d. Doubling to $M=8$ IPUs halves the per-core load and admits $G_\mathrm{src}=64$.

Even though grouping determines how synapses are distributed, it does not change the mean synaptic work per thread. Each thread is predicted to process on average $2.7$ active synapses per time step for all group configurations (Sec.~\ref{sec:methods_prediction_workload}, Fig.~\ref{fig:performance}e). Scaling groups nonetheless impacts performance. With more source groups, a thread receives fewer spikes, but each spike activates more synapses on it, so that the
predicted max-to-mean ratio of active synapses rises from roughly $5$ at $G_\mathrm{src}=4$ to $14$ at $G_\mathrm{src}=92$, compared with $4.4$ for an ungrouped distribution (Fig.~\ref{fig:performance}f). Conversely, the number of spikes each thread has to look up decreases with $G_\mathrm{src}$. At worst, a thread scans about $4$ spikes per step at $G_\mathrm{src}=16$, compared with $10.4$ in the ungrouped case even with ideal target filtering (see Tab.~\ref{tab:montecarlo}). Together with the reduction cost, which grows with the target group size and is similar to the memory scaling, this trade-off between lookup work and synaptic load imbalance favours intermediate group
configurations for performance (Fig.~\ref{fig:performance}b).

\subsection{Performance}
\label{sec:performance_results}
\begin{figure}
    \centering
    \includegraphics[width=0.38\textwidth]{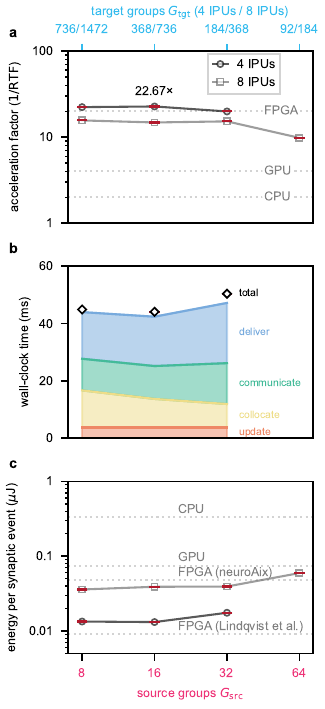}
    \includegraphics[width=0.61\textwidth]{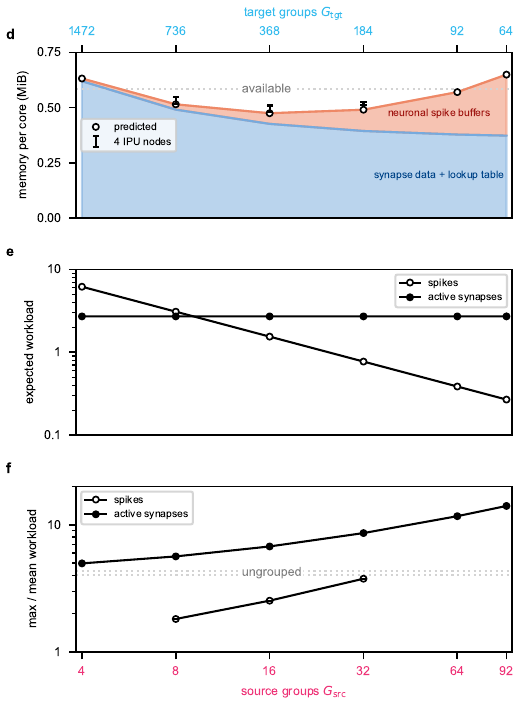}
    \caption{\textbf{Performance, memory and spike statistics across group configurations.} All plots show data for different numbers of source groups $G_\mathrm{src}$ on the bottom axis, whereas top is the corresponding number of target groups $G_\mathrm{tgt}$. Whenever applicable, error bars are shown for three differing seeds run for $T_\mathrm{bio}=\SI{15}{\minute}$ each. \textbf{a.}~shows speedup factors ($1/\mathrm{RTF}$); the horizontal lines show the best-performing setup for different types of hardware, on both $M=4$ and $8$ IPU configurations. Top to bottom: FPGA cluster~\cite{kauth2023neuroaix}, GPU~\cite{gpuRuntimeConstr} and CPU~\cite{Kurth_2022}. \textbf{b.}~Breakdown of simulation time into algorithm phases on four IPUs for $T_\mathrm{bio}=\SI{1}{\second}$: \emph{Update}, \emph{Collocate}, \emph{Communicate} and \emph{Deliver}. The marker indicates total simulation time. \textbf{c.}~shows the energy per synaptic event (see Sec.~\ref{sec:method_energy}) for $4$ and $8$ IPU configurations. For FPGAs, the level of measurement (e.g. board, system, device) differs depending on available data and tooling (see Sec.~\ref{sec:results_energy}).  \textbf{d.}~Per-core memory usage with stacked contributions of different data structures (see legend). Available memory per core on the reference system is indicated by the horizontal dotted grey line. For prediction (open circle) see Sec.~\ref{sec:methods_prediction_mem_usage}. \textbf{e.}~Expected spikes per source group (open) and active synapses per thread (filled) per time step. \textbf{f.}~Max-to-mean ratio of active synapses per thread (filled, Monte Carlo) and of spikes
per source group (open, measured). Dotted line: ungrouped placement ($G_\mathrm{src}=1$).
Values in Table~\ref{tab:montecarlo}.}
    \label{fig:performance}
\end{figure}

\begin{table}[!htbp]
    \centering
\caption{Performance (RTF) and energy-efficiency ($E_\mathrm{syn}$) benchmark results for full-scale PD14 simulations, ordered by publication year and showing only the best reported performance or energy-efficiency results for each architecture. Energy values depend on the measurements reported in the original studies and therefore may be device-, board-, or system-level. For the nearest competitors, footnotes specify which hardware components are included in the power measurements.
}
    \label{tab:competition}
\setlength{\tabcolsep}{1pt}
\begin{tabular}{cccccc}
    \toprule
    \textbf{RTF} & \textbf{Simulator} & \textbf{Ref.} & \textbf{$E_{\mathrm{syn}}$ (\si{\micro\joule})} & \textbf{\# Nodes} & \textbf{System}\\
    \midrule
    1.00  & SpiNNaker 1~\cite{spinnakerRealtime}   & Rhodes et al. (2019)     & 0.601 & 12  & 48x18xARM-968        \\
    0.53  & NEST CPU~\cite{Kurth_2022}      & Kurth et al. (2022)      & 0.48  & 2   & 2 AMD EPYC Rome 7702    \\
    0.27  & GeNN~\cite{gpuRuntimeConstr}          & Golosio et al. (2023)    & 0.074 & 1   & RTX 4090           \\
    0.05  & neuroAIx~\cite{kauth2023neuroaix}      & Kauth et al. (2023)      & 0.048\textsuperscript{†} & 35  & NetFPGA SUME           \\
        0.79 & horiz/mono/s~\cite{lindqvist2024fast} & Lindqvist et al. (2024) & 0.021 / 0.009\textsuperscript{‡} & 1 & 1 Agilex 7 FPGA  \\
\textbf{0.044} & \textbf{IPU} & \textbf{This work} & \textbf{0.013} & \textbf{4} & \textbf{Graphcore Bow IPU}  \\
    \bottomrule
\end{tabular}
\par\smallskip
{\footnotesize
\textsuperscript{†} Board-level power, including external memory and peripherals.\\
\textsuperscript{‡} Device power only, excludes host and system overhead.
}
\end{table}
We achieve state-of-the-art acceleration of $22.7\times$ over biological real-time, $1.11\times$ faster than the previous record~\cite{kauth2023neuroaix} (Fig.~\ref{fig:performance}a). This is done on $M=4$ IPUs, the smallest number of nodes that can fit the full-scale microcircuit in memory, and with $G_\mathrm{src}=16$ due to the load balancing and memory trade-offs discussed in Sec.~\ref{sec:memory_usage_workload}. Doubling to $M=8$ IPUs lowers the speedup to at most $15.6\times$. The benchmarked microcircuit model is computationally lightweight enough that it above all benefits from locality, with neuron updates being the smallest contributor to simulation time. The error bars in Fig.~\ref{fig:performance}a show minimal standard deviation due to the static graph compilation on the IPU and semi-fixed connectivity scheme of the microcircuit (Sec.~\ref{sec:PD14}).

\DeclareSIUnit\Joule{Joule}

\subsection{Energy}
\label{sec:results_energy}
For the fastest configuration, three runs with different connectivity seeds consume $\SI{11361.77\pm5.34}{\Joule}$ of device energy per run and produce an average of $\num{222580247}$ spikes over $T_\mathrm{bio}=\SI{15}{\minute}$. Using the mean neuronal fanout $K_N=3880$, we obtain $\SI{13.16}{\nano\joule}$ per synaptic event, where one event denotes the delivery of one emitted spike to one post-synaptic target. At the device level, this is competitive with the record $\SI{9}{\nano\joule}$ FPGA-fabric-only estimate reported by Lindqvist and Podobas~\cite{lindqvist2024fast}. While energy usage is $3.6\times$ lower than neuroAIx’s reported $\SI{47.55}{\nano\joule}$, neuroAIx’s measurement includes external memory and board peripherals, while ours covers aggregate per-IPU device power. This ratio therefore compares reported values with different measurement boundaries. Published results across architectures are compared in Tab.~\ref{tab:competition} and Fig.~\ref{fig:performance}c. An extended version of the table, including process node and external-drive configuration, is provided in Tab.~\ref{tab:competition_full}.

\subsection{Build and Compilation}
For the full-scale microcircuit on $M=4$ IPUs, most of the build time is spent generating the lookup table used for spike routing, which takes $\SI{129.92\pm2.54}{\second}$. The lookup table records how many local targets each incoming source index has for a given source-target group pair. Its construction requires local sorting of target indices to avoid threads accessing the same ring buffer slot during routing, along with alignment steps to account for padding. The Graphcore software stack also requires an internal graph compilation step, which takes $\SI{4.21\pm1.03}{\second}$ for the same samples. Once built, the computational graph can be stored and loaded from file in subsequent runs, allowing parameters that are not tied to tensor sizing or logic flow, such as simulation duration, to change without repeating these build steps.

\section{Methods}
\label{methods}
\subsection{Benchmark Network Model}
\label{sec:PD14}
Our primary benchmark target is the well-established microcircuit model by Potjans and Diesmann~\cite{potjans2014}. The model represents \SI{1}{\milli\meter\squared} of cortical tissue incorporating experimental connectivity data from rat and cat neocortex, spanning around 300 million synapses. It is subdivided into four layers, each with both excitatory and inhibitory neuron populations. Lateral connectivity is random, meaning there is no inherent locality that can be exploited during spike propagation. Synapse data, meaning primarily weights and delays, are drawn from population-specific distributions. Neuron pairs (pre, post) are randomly picked within population pairs up to a fixed total number derived from the inter-population probabilities. The model is considered a building block of cortical function and features as a basic unit in larger models~\cite{Schuecker17,Schmidt18_2,senkMesocircuit}. 

Due to these characteristics, the microcircuit model has become a stand-in for measuring neural simulator efficiency across platforms and software packages. Senk et al.~\cite{Senk_2026} define a recipe for benchmarking and validation, which we make use of. The performed tests, comparing population mean rates, correlation, and inter-spike intervals, create fundamental computational requirements in maintaining the desired accuracy. Common standards to ensure this are:

\begin{itemize}
    \item \textbf{Individual representation of synapse data}, which imposes a minimum memory requirement. This includes the synapse target (connectivity), weights, and delays. Although previous work has established that single-float (32-bit) precision weights still produce statistically correct dynamics~\cite{knight2018}, this per-synapse data still dominates the overall memory requirement for models. Especially on accelerators and other memory-constrained devices, this limits the options for parallel algorithms that would require duplicated data structures per thread. Procedural connectivity generation can reduce this footprint by drawing connections on-the-fly from seeded distributions~\cite{knight2021connGen}, allowing reproducible simulations of larger static networks at competitive performance. Here, we use explicit synapse data to follow the full-density benchmark specification and to keep the routing and memory requirements directly comparable across simulator backends.

    \item \textbf{Strictly temporally sequential simulation} with a $\Delta t$ of \SI{0.1}{\milli\second}. This significantly restricts the room for parallelisation, as the computational burden of updating \emph{exactly integrated}~\cite{rotter1999exact} current-based (CUBA) LIF neurons is already rather small. The bottleneck of delivery and communication oftentimes makes scaling up the number of processes naively more costly, due to more disjoint memory and physical separation, outweighing gains from update parallelism.
\end{itemize}
Given these requirements, frequent memory access to look up and route incoming spikes and the corresponding synapse data (deliver phase) is unavoidable. On conventional hardware, this requires carefully chosen data layouts to ensure predictable access to allow pre-fetching and minimise cache misses~\cite{pronold2022routing}. Even with these optimisations applied, delivery time takes up a significant portion of simulation time on a microcircuit scale. We simulate the full-scale PD14 model at a $\Delta t=\SI{0.1}{\milli\second}$ simulation step: $N=\num{77169}$ neurons, approximately $\num{3e8}$ synapses ($K_N=\num{3880}$ per neuron on average), mean firing rate $\nu_N=\SI{3.2}{\hertz}$, and synaptic delays starting at $d_\mathrm{min}=\SI{0.1}{\milli\second}$. Unless stated otherwise, runs use $M=4$ Bow IPUs (one Bow-2000), providing $C=M \cdot C_M=\num{5888}$ ($C_M=1472$ per IPU) cores with $T_C=6$ hardware threads and \SI{624}{\kibi\byte} local SRAM per core~\cite{graphcore_bow2000}.

\subsection{Reference Distributed Memory System}
\label{sec:IPU}
Under the hood, Graphcore's software stack implements a bulk synchronisation pattern~\cite{valiant1990bsp}. This scheme divides computation into a local update step, a communication phase where cores can send and request non-local data, and finally a global synchronisation barrier. Random access to core-local SRAM comes at a constant cost of 6 cycles. To hide this instruction latency, the IPU employs a barrel processor with 6 threads, where each is scheduled in round-robin fashion each cycle. Every core is capable of MIMD, meaning every thread can execute a different context.

Any IPU program is compiled into a static graph, where codelets (akin to CUDA kernels) represent vertices and edges are data dependencies. The implications of these paradigms are that programs must be reasonably well load balanced and that data, represented by tensors at runtime, is of static shape. The latter requirement creates an algorithmic challenge for efficient implementation of structures that are usually represented dynamically in neural simulations, such as spike buffers.
IPUs support single (32-bit) and half (16-bit) precision floating point arithmetic. The software stack extensively makes use of low and mixed precision operations and offers additional utility to alleviate precision loss, for instance via stochastic rounding~\cite{forsythe1959reprint}.

The IPU however, both in terms of hardware and software stack, primarily targets machine learning use-cases. Previous work has used it to train and run spiking neural networks in a deep-learning setting~\cite{sun2024exploiting, finkbeiner2024harnessing}, whereas our simulation of biologically constrained networks with explicit synapses and delays relies on fine-grained control over data layout and movement. Fortunately, Graphcore provides both a high-level Python front-end and a lower level C++ ecosystem, \texttt{poplar}, which largely fulfils these requirements. To achieve peak performance, we nonetheless rely on optimised handcrafted codelets and algorithms to cover our specific simulation steps and work around certain IPU design choices, further described in Sec.~\ref{sec:indepthalgo}.

\subsection{Algorithm Implementation}
\label{sec:indepthalgo}
We implement the four simulation phases introduced in Sec.~\ref{sec:algooutline} (and visualised in Fig.~\ref{fig:overview}) as a static \texttt{poplar} computational graph using custom C++ codelets. The following subsections describe the data layouts and operations used in each phase. During graph construction, a layout checker verifies that tensors intended to remain core- or group-local are not accessed outside their assigned scope. The neuron state update step of the algorithm is straightforward and fully described by the outline given in Sec.~\ref{sec:algooutline}.

\subsubsection{Collocate}
\label{sec:indepthagg}
Collocation is split into local on-core and source group-wide neuron aggregation steps. Aggregation steps can be viewed as a parallel prefix-sum algorithm~\cite{hillis1986data} with a fixed number of stages. Indices of spiking neurons are recorded in integer precision to comply with the minimum word size and to leverage the control-flow-optimised execution pipeline.

\textbf{Core-local Aggregation.} First, a single thread per core gathers the source group local index $n_j^\mathrm{src}$ of all $N_C$ locally hosted neurons that crossed the firing threshold into a local buffer sized $b_\mathrm{core}$. The expected number of spikes per time step and per core for a given model with an average neuronal firing rate of $\nu_N$ is specified by $\nu_N N_C \Delta t$. For the microcircuit, for most realistic group combinations, this value falls below one. The buffer must be sized to accommodate the maximum per-time step spike volume. Load balancing measures described in Appendix~\ref{app:loadbalance} help minimise this peak capacity requirement.

\textbf{Source Group Aggregation.} In the next stage, all $G_\mathrm{tgt}b_\mathrm{core}$ potential spike source indices are aggregated into a combined source group-wide buffer of size $b_\mathrm{src}$. For this, we use all available threads on $c_\mathrm{agg}$ cores in the centre of the group (assuming linear layout), each writing into fixed offsets $b_\mathrm{src}/T_C$ within the output. Every thread denotes the number of elements found $s_i$ in the $0\text{th}$ slot of its allocation.

\textbf{Buffer Packing.} Finally, we run another multithreaded codelet on the same aggregation cores to write the indices into a contiguous buffer for all grouped aggregation cores $c_\mathrm{agg}$. This is referred to as the send buffer and forms the actual payload we will communicate to targets in the next step. To do so, every thread calculates the cumulative sum of spikes up to its own ID, defined as $\sum_{i=0}^{\mathrm{t-1}} s_i$, and then uses the resulting offset into the final output buffer to copy over indices. A count of all spikes in the group aggregation, $n_\mathrm{count}$, is written to index $0$. Viewed across all groups, this results in a buffer of shape $(G_\mathrm{src}, c_\mathrm{agg}, n_\mathrm{final})$ where $n_\mathrm{final}$ is the upper limit for the count of spikes per group and each contains $N^{\mathrm{src}}_\mathrm{inc} \subset N_\mathrm{src}$ source group indices.

\subsubsection{Communicate}
The \emph{communicate} phase transfers the packed send buffers produced during the collocate phase from source groups to target-group cores. Within each target group, a core with $k_j^\mathrm{tgt}$, $j\in\{0,\dots,G_\mathrm{src}-1\}$, receives the send buffers of the corresponding source group $\mathcal{G}_{\mathrm{src},j}$. The target group collectively receives the buffers of all source groups, while each of its cores receives data from only one source group.
On the IPU, this exchange is specified implicitly. Target-side access to a source-group send buffer establishes the required data dependency, and \texttt{poplar} performs the data movement before the \emph{deliver} computation.

\subsubsection{Deliver}
The \emph{deliver} phase first maps the source group local neuron indices contained in the received buffers to locally stored synapses and accumulates the weighted events in the target neurons' ring buffers. Each target-group core stores the synapse data originating from its corresponding source group and maintains a private ring buffer contribution covering all neurons in the target group. In a second step, the duplicate ring buffer views are reduced to ensure each neuron has a full view of all exchanged spikes in a time step. 

\textbf{Spike Delivery.} Translation of the received buffers' source group local indices $N^{\mathrm{src}}_\mathrm{inc}$ to synapse data as well as parallelisation of the whole process is enabled by the lookup table. During build time, we create it as a tensor of shape $(G_\mathrm{tgt}, G_\mathrm{src}, N_\mathrm{src}, T_C + 1)$. The first two dimensions identify a unique target group core by the source-target group pair, the third spans the maximum number of incoming source group local indices and the last splits the table by threads. It thus defines a cumulative range of indices into local synapse data. This thread split must account for neuron boundaries, as splitting the same synapse targets across threads could lead to race conditions. Depending on our choice of $(G_\mathrm{src},G_\mathrm{tgt})$, the lookup table's maximum range value for local synapse data will be small enough to encode it as 16-bit integers. The $T_C$ dimension partitions the local target neurons $N_\mathrm{tgt}$ into contiguous blocks, one per thread, rather than partitioning synapses directly. Due to the incoming send buffers' shape, $(G_\mathrm{src}, c_\mathrm{agg}, n_\mathrm{final})$, an outer and an inner loop are required for delivery of all spikes. The effective size of the group aggregation dimension $n_\mathrm{count}$ is stored at index $0$, and serves as the limit for the inner loop to avoid iterating over empty buffer slots. Finally, we can use the index range retrieved from the lookup table to store synapse weights at position $(t_\mathrm{now} + d)\mod d_{\max}$ in the ring buffer, where $t_\mathrm{now}$ is a running index in range $[0, d_{\max})$ signalling the current time step and $d$ is the synapse delay used to offset. In the microcircuit model, the synaptic time constant $\tau_\mathrm{syn}$ is the same for excitatory and inhibitory connections, which allows us to merge both types of possible weights into one dimension in the ring buffer.

\textbf{Reduce.} Each target group core handles a unique ($\mathcal{G}_{\mathrm{src},i}, \mathcal{G}_{\mathrm{tgt},j}$) combination in terms of spike routing. However, since one source group's send buffer is sent to every target group, the buffers inevitably contain source group indices matching neuron targets beyond what is local on any one core. Conversely, one target group core only handles spikes from one source group, but must be able to receive spikes from all.
For this reason, every core must be able to accept spikes with targets located on any other core within its target group. This is reflected in the shape of the ring buffer $(C, d_{\max}, N_\mathrm{tgt})$, which must be reduced along the core dimension for $d=t_\mathrm{now}$ each time step. The $G_\mathrm{src}$ per-core copies of shape $(d_{\max}, N_\mathrm{tgt})$ are summed element-wise at slot $t_\mathrm{now}$, yielding one input value per target neuron. The operation is implemented using a \texttt{poplar} library method which leverages manual assembly codelets. Since group sizes are aligned with the IPU node boundary, this reduce operation is always IPU-local and therefore incurs no global communication.

\subsection{Modelling Choices}
In the standardised version of the microcircuit model\footnote{\url{https://github.com/INM-6/microcircuit-PD14-model/}}, synaptic delays are drawn from a truncated normal distribution starting at the minimum delay and unbounded above $[d_\mathrm{min}, \infty)$. When building the computational graph and compiling the codelets on the IPU, only fixed-size data structures are allowed. As the maximum delay $d_{\max}$ impacts the sizing of the ring buffer $(C, d_{\max}, N_\mathrm{tgt})$, the maximum delay has a strong effect on the total memory usage. No delays are truncated in our implementation, but $d_{\max}$ directly scales the ring buffer memory
(Sec.~\ref{sec:methods_prediction_mem_usage}). Bounding the delay distribution offers a way to fit models slightly exceeding the available memory, at the cost of a small deviation from the model's specification. Synaptic weights could also be stored in half precision, which halves their memory footprint and passes the statistical validation (Appendix~\ref{app:verification}). However, since the routing codelet processes spikes scalarly and the IPU's minimum word size is 32 bits, the required realignment cancels out the speedup of the reduction, so all reported results use full precision.

\subsection{Codelet Optimisation}\label{sec:vec}
Graphcore's codelet compiler \texttt{popc} is based on LLVM~\cite{llvm2004}, which allows for interpretation and iterative optimisation of performance-critical code by compiling to an intermediate representation and checking the issued instructions. IPU memory is 64 bits wide, with a 2-way interleaved region that additionally supports 128-bit loads and stores~\cite{graphcore_ipu21_isa}, and we vectorise wherever the data layout permits. All input and output data is defined as tensors via \texttt{poplar} and mapped to a one-dimensional naturally aligned vector in memory for processing in codelets. The neuron update processes neuron pairs with 64-bit \texttt{float2} arithmetic, and the reduce step reads and clears the current ring buffer slice with 128-bit \texttt{float4} accesses before summing it across source groups with \texttt{poplar}'s vectorised reduction. Aggregation and routing codelets benefit less. Spike indices are integers, and the IPU's integer pipeline lacks vector instructions~\cite{graphcore_ipu21_isa}: the aggregation step can scan its input in \texttt{int4} blocks to skip empty entries, but compacting the remaining spikes is scalar. Routing performs data-dependent scatter-adds into the ring buffer, indexed by target neuron and delay, and therefore remains scalar regardless of data type. Spike delivery is consequently the most expensive phase (Fig.~\ref{fig:performance}b). Since the scatter-add cannot be vectorised, its cost per step is set by the number of active synapses on the most loaded thread, which motivates the workload analysis in Sec.~\ref{sec:methods_prediction_workload}.

\subsection{Benchmarking and Analysis}
To assess whether the benchmark model can be simulated on the reference system and to characterise the expected workload and imbalance, we provide analytical predictions of the per-core memory footprint and the per-thread computational load. The analysis is supported by Monte-Carlo estimates of the load imbalance as well as measured data. The predictions are evaluated as a function of the source group count (see Fig.~\ref{fig:performance}), which controls how neurons and their synapses are partitioned across cores (Sec.~\ref{sec:indepthalgo}).
\subsubsection{Setup}
The reference system (Sec.~\ref{sec:IPU}) comprises $M = 4$ IPU nodes, each with $C_M = 1472$ cores and $T_C = 6$ hardware threads per core, giving a total of $C = M\,C_M = 5888$ cores and $T = C\,T_C = \num{35328}$ threads. The benchmark network is a cortical microcircuit model (Sec.~\ref{sec:PD14}), with $N = \num{77169}$ neurons and on average $K_N = 3880$ incoming (equivalently outgoing) synapses per neuron. We assume a mean firing rate of $\nu_N=\SI{3.2}{\hertz}$ and a simulation resolution of $\Delta t = \SI{0.1}{\milli\second}$. The minimum synaptic delay equals the simulation step size, which means that spikes need to be communicated every step. The predictions do not take into account the different neuronal populations and their specific firing rates and connectivity.

\DeclareSIUnit\Byte{Byte}

\subsubsection{Memory Usage}
\textbf{Prediction.}\label{sec:methods_prediction_mem_usage}
On average $N_C = \lceil N / C\rceil$ neurons and $K_C = N_C\,K_N$ synapses are placed on each core. The cores are partitioned into $G_\mathrm{src}$ source groups and $G_\mathrm{tgt} = C / G_\mathrm{src}$ target groups. We consider $G_\mathrm{src} \in \{4, 8, 16, 32, 64, 92\}$. A source group comprises $N_{\mathrm{src}} = G_\mathrm{tgt}N_C$ neurons and a target group holds $N_{\mathrm{tgt}} = G_\mathrm{src}N_C$ neurons.
For the prediction of the per-core memory usage, we take into account the contributions of synapse data (weights, delays, targets), the lookup table storing synapse-data positions, and the neuronal spike buffer (Fig.~\ref{fig:performance}d). We assume that each synapse takes up $m_\mathrm{syn}= \SI{7}{\Byte}$ of memory, which is composed of a $m_\mathrm{w} = \SI{4}{\byte}$ weight, a $\SI{1}{\byte}$ delay, and a $\SI{2}{\byte}$ target group local neuron ID. The lookup table, viewed per-core, has two dimensions and holds $T_C + 1$ local connection indices of $m_{\mathrm{lcid}} = \SI{2}{\byte}$ for each neuron of the sending source group. Altogether, synapse data and lookup table require
\begin{equation}
\mathcal{M}_{\mathrm{syn}} = N_{\mathrm{src}}\,(T_C + 1)\,m_{\mathrm{lcid}}
    + K_C\,m_{\mathrm{syn}}.
\end{equation}
The two-dimensional neuronal spike buffer reserves one weight-sized slot per target group local neuron and per potential delay step, where we assume a maximum of $d_{\max} = 60$ steps. This yields a memory usage of
\begin{equation}
\mathcal{M}_{\mathrm{buf}} = N_{\mathrm{tgt}}\,d_{\max}\,m_{\mathrm{w}}
\end{equation}
for the neuronal spike buffers. In total, the predicted per-core memory amounts to
\begin{equation}
\mathcal{M} = \mathcal{M}_{\mathrm{syn}} + \mathcal{M}_{\mathrm{buf}},
\end{equation}
which we compare against the available budget of approximately \SI{624}{\kibi\byte} (Fig.~\ref{fig:performance}d).

\textbf{Measurement.} The actual memory measurements include a few quantities omitted from the prediction because they are difficult to predict ahead of time. This includes, for instance, code storage, live exchange buffers (created automatically when data is communicated) and padding for alignment purposes. These contributions are relatively minor, which is why the error bars in Fig.~\ref{fig:performance}d are shown to line up closely with the prediction. Memory is recorded using the \texttt{PopVision Analysis Library (libpva)}, with which we programmatically read the mean memory of $G_\mathrm{src}$ cores (one aggregation core per group). We then take the mean and standard deviation across three connectivity seeds.

\subsubsection{Workload}
\textbf{Prediction.} \label{sec:methods_prediction_workload}
Each source group has target synapses on one core of each target group. Hence, the synapses of one source group are distributed over
\begin{equation}
T_{\mathrm{tgt}} = G_\mathrm{tgt}T_C
\end{equation}
destination threads. The probability for a neuron to spike in a time step is defined as
\begin{equation}
p = \nu_N\, \Delta t .
\end{equation}

The expected spike count per source group, as shown in Fig.~\ref{fig:performance}e, then is 
\begin{equation}
    S_\mathrm{src} = \frac{Np}{G_\mathrm{src}}.
\end{equation}
Increasing the number of source groups reduces the expected number of spikes per source group, $S_\mathrm{src}$, but concentrates each neuron's $K_N$ synapses on fewer cores, since a source group spans only $G_\mathrm{tgt}=C/G_\mathrm{src}$ cores. Both effects cancel out exactly, so that the expected number of active synapses per thread and step can be written as:
\begin{equation}
A = S_\mathrm{src}\,\frac{K_N}{T_\mathrm{tgt}}
=\frac{Np}{G_\mathrm{src}}\cdot\frac{K_N\,G_\mathrm{src}}{T} = \frac{N K_N\,p}{T} \approx 2.7
\end{equation}
and is independent of the group configuration (see Fig.~\ref{fig:performance}e). Each thread also scans all $S_\mathrm{src}$ spikes of its source group in the \emph{deliver} loop, so the lookup work per thread decreases with $G_\mathrm{src}$.
Since the IPU's cores must sync after each operation (see Sec.~\ref{sec:IPU}), a simulation step is only as fast as its slowest parallel operation. So the relevant 
performance bottleneck is the expected maximum load relative to the mean: $\mathbb{E}[\max_\mathrm{thread} A_\mathrm{thread}]/A$. Although the mean number of active synapses per thread is constant, a larger $G_\mathrm{src}$ means that each target thread receives fewer spikes (fewer cores per corresponding source group), but each spike activates more synapses on it (more cores in the target group it belongs to). The per-thread load therefore fluctuates more strongly, and its maximum over all threads grows, as shown in Fig.~\ref{fig:performance}f. The figure's ratio of active synapses $\mathbb{E}[\max_\mathrm{thread} A_\mathrm{thread}]/A$ is estimated by Monte Carlo simulation over 100 iterations of $\num{1000}$ time steps each, drawing each neuron's $K_N$
target threads uniformly from the $T_\mathrm{tgt}$ threads of its source group and its spikes independently with probability $p$. As a reference, the same procedure is applied to an ungrouped distribution in which synapses are drawn from all $T$ threads ($G_\mathrm{src}=1$).

\textbf{Measurement.} The analytical solutions as well as Monte Carlo simulations make some idealised assumptions, such as a homogeneous, uncorrelated firing rate and a uniform split of synapses over target threads. In our simulation code, in order to avoid overlapping write access, each target core thread $k \in \{0,\dots,T_C-1\}$ is assigned a fixed slice of possible local neuron IDs:
\begin{equation}
   n_{k,\mathrm{start}} = k\left\lceil\frac{G_\mathrm{src}N_C} {T_C}\right\rceil.
\end{equation}
Since neurons of each population are interleaved across source groups, each block contains a mix of populations, so the impact of this difference is expected to be small. Empirical spike counts per source group (and the resulting max-mean ratios) are obtained from recorded spike trains (spike output enabled, untimed runs of $\SI{15}{\minute}$ biological time, 3 seeds), binned per time step and source group. Per-thread active synapse counts were not manually recorded.

\subsubsection{Simulation Time}
\label{sec:method_performance}
All results are measured by recording cycle counts using IPU native methods over the whole simulation duration. Cycle counts were converted to wall-clock time using the IPU device clock frequency $f=\SI{1.85}{\giga\hertz}$. Real-time factors are reported as $\mathrm{RTF}=T_\mathrm{wall}/T_\mathrm{bio}$, where values below one indicate faster-than-real-time simulation. The primary codelets involved in a simulation cycle are additionally timed to allow for a time breakdown. Performance measurements generally exclude copying spikes to host. Spike-train output is enabled only for (untimed) validation runs, as is standard practice for acceleration comparison~\cite{Senk_2026}. Static buffer sizes (e.g.\ $b_\mathrm{core}$, $n_\mathrm{final}$) and the aggregation granularity $c_\mathrm{agg}$ were tuned per configuration by grid search. Each configuration was first run in a debug build that checks all aggregation buffers for overflow and aborts if one occurs. Timed runs use the same seeds and durations with these checks disabled.

\subsubsection{Energy}
\label{sec:method_energy}
Energy usage was measured using the IPU's built-in \texttt{gcipuinfo} command, which reads out the momentary per-IPU power consumption. In order to periodically record values during a simulation we use the JSC power tool\footnote{\url{https://github.com/FZJ-JSC/jpwr}} described in~\cite{jpwrLore}, which spawns a monitoring thread on the host and aggregates the power measurements. Running this over 3 seeds for full-scale simulations of $T_\mathrm{bio}=\SI{15}{\minute}$ biological time for the fastest configuration and querying power every $\Delta t_\mathrm{sample}=\SI{0.1}{\second}$ of runtime, we can calculate total energy $E_r$ consumed by integrating per-IPU power $P$ over IPUs $k$ and individual samples $i$ for one run $r$:
\begin{equation}
E_r =
\Delta t_{\mathrm{sample}}
\sum_{i=1}^{n_r}
\sum_{k=1}^{M}
P_{r,k,i}.
\end{equation}
Based on standard practice~\cite{SHARP2012110,spinnakerpd14}, the number of synaptic events was estimated from the number of emitted spikes $S_r$ and the mean neuronal fanout $K_N$, with which we calculate the energy per synaptic event:
\begin{equation}
E_{\mathrm{syn},r}=\frac{E_r}{S_r K_N}.
\end{equation}
Reported values are the mean and standard deviation of $E_{\mathrm{syn},r}$ across the three runs.

\subsubsection{Build Time}
\label{sec:method_build}
Simulation parameters and model data are specified through a Python frontend. Synapse and neuron inputs are provided in a structure of arrays (SoA) format, re-organised for the chosen C++ layout, and converted to runtime tensors. As a result, changes to hyperparameters or input models do not require C++ re-compilation. This keeps the edit-run cycle short compared with static simulators that achieve comparable simulation speedups only after substantially longer compilation steps, such as FPGA-based systems~\cite{kauth2023neuroaix,heittmann2022simulating}. Because the model implementation is decoupled from the hardware-specific graph construction, new synapse or neuron models only require software re-compilation, adding seconds of build time at most. Build times were recorded for the IPU's host system, featuring an \textit{AMD EPYC 7742} processor.

\section{Discussion}
\label{sec:discussion}
In this work, we demonstrate that distributed on-chip memory can move spiking network simulation beyond the performance envelope of conventional many-core systems. Our group-based routing strategy maps spike exchange onto structured source-target communication, reducing runtime filtering while preserving the flexibility of a programmable accelerator. Applied to the full-scale cortical microcircuit, the resulting IPU simulator reaches $22.7\times$ biological real-time, validates against NEST results, and combines high speed with short build times and competitive energy efficiency. These results establish distributed-memory hardware as a compelling platform for large-scale computational neuroscience, and provide an algorithmic template for exploiting locality in future neural simulators.

While the microcircuit achieves peak performance at just $M=4$ IPUs, this can mainly be attributed to the low computational load, which makes a compact fit in memory and short communication pathways crucial. Since on-chip memory grows linearly with the number of IPUs (about \SI{0.9}{\giga\byte} each~\cite{graphcore_bow2000}), larger systems provide tens of gigabytes of SRAM. This capacity may benefit models that make more use of update parallelism or have spatially constrained activity, for which the additional communication cost of spreading across more nodes, as seen here when moving from $M=4$ to $M=8$, is less dominant.

When respecting the modelling requirement that synapse data must be explicitly stored on the device, the next largest benchmark models, such as the mesocircuit~\cite{senkMesocircuit} or the multi-area model~\cite{schmidt2018multi}, are at least an order of magnitude larger in neuron count and therefore hard to fit into purely local memory. This is a capacity limit of the present explicit-connectivity implementation rather than a limitation of the group-based routing strategy itself. For long-running studies with evolving connectivity, such as structural plasticity experiments, the IPU is nevertheless a promising target: its achieved speedup directly reduces the wall-clock cost of biologically long simulations, while hardware-supported random number generation~\cite{hanlon2022fast} can make stochastic target selection and connectivity updates comparatively cheap. The IPU's static tensor model means that such dynamics would still need bounded data structures, for example through over-allocation or plasticity rules with explicit upper limits, but these constraints are compatible with a broad class of structural plasticity experiments. At $22.7\times$ real time, a full day of biological time of the microcircuit is simulated in about an hour on $M=4$ IPUs. This makes the platform a natural candidate for plastic networks, whose biologically relevant timescales of days to weeks remain prohibitive for conventional simulators.

\backmatter

\section*{Acknowledgements}
The presented conceptual and algorithmic work is part of our long-term collaborative project to provide the technology for neural systems simulations\footnote{\url{https://www.nest-initiative.org}}.
This research used resources of the Argonne Leadership Computing Facility, a U.S. Department of Energy (DOE) Office of Science user facility at Argonne National Laboratory and is based on research supported by the U.S. DOE Office of Science-Advanced Scientific Computing Research Program, under Contract No. DE-AC02-06CH11357.
This work is partly funded by Volkswagen Foundation - 0071143 (STRUCTICITY) and the German Research Foundation (DFG) - 545776403/FOR5880 (Mod4Comp). We would further like to thank Jette Oberländer for her help in reviewing this manuscript.

\section*{Competing interests}
The authors declare no competing interests.

\bibliography{bibliography}

\bigskip

\begin{appendices}

\section{Verification of implementation}
\label{app:verification}
\begin{figure}
    \centering
    \includegraphics[width=1.0\linewidth]{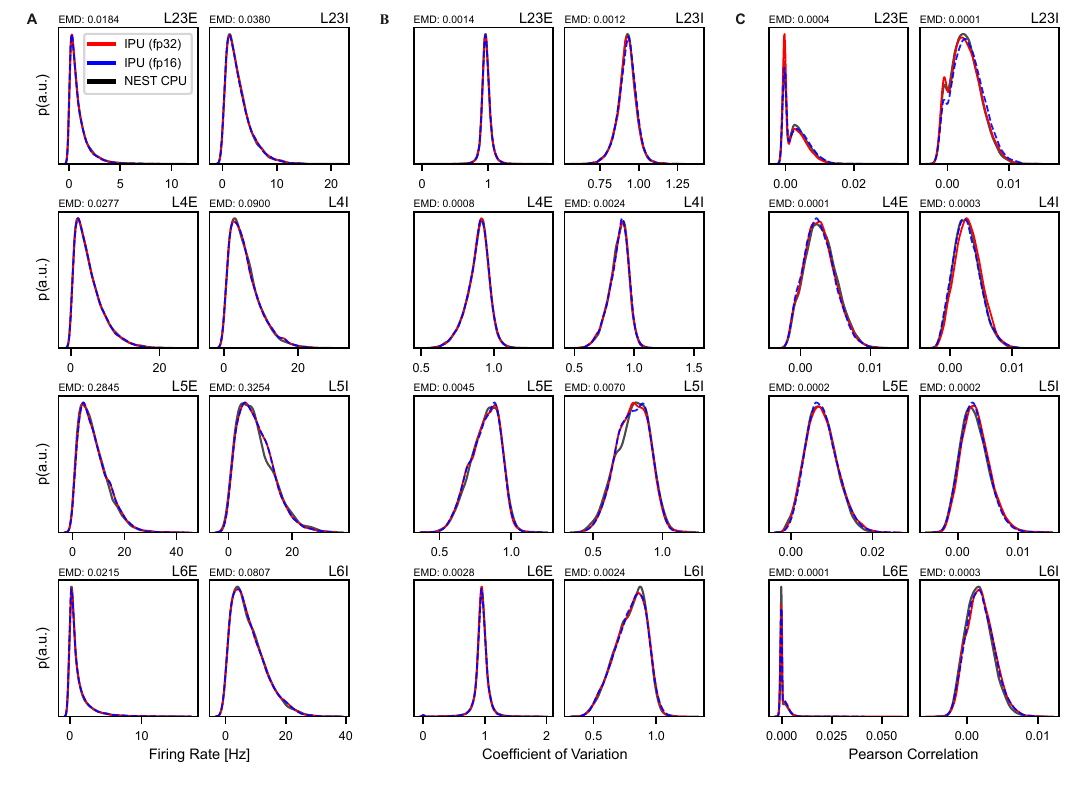}
    \caption{Probability density plot of the three metrics (\textbf{A:} firing rates, \textbf{B:} coefficient of variation and \textbf{C:} Pearson correlation) chosen to validate simulation results plotted for every cortical layer and excitatory/inhibitory population. The baseline run from an established CPU simulator, NEST, is shown in black along with our data for both full- (red) and half (blue, dotted) precision. Earth Mover's Distance (EMD) between full precision IPU and NEST is labelled in the top right for each population.}
    \label{fig:verification}
\end{figure}

We verify the correctness of spike trains by applying the established statistical tests described in~\cite{Senk_2026} to the output of a $\SI{15}{\minute}$ biological time full-scale simulation after a five second warm-up period, during which spikes are discarded. We generate the microcircuit connectivity for a random seed on the IPU host side and plot the probability distributions of the resulting metrics in comparison to a random seed run performed on the CPU version of \texttt{NEST 3.10}~\cite{schofmann_2026_18268837}, shown in Fig.~\ref{fig:verification}. Although we do not use 16-bit floating point weights for our simulation results, we include half precision with stochastic rounding to show that it is a viable choice for neural simulations on accelerators with dedicated support for 16-bit arithmetic. We also include the Earth Mover's Distance (EMD) between the full precision IPU run and the NEST baseline.

Per population and their respective neurons, we plot the \textbf{firing rate} over the whole simulation duration, the \textbf{coefficient of variation} between spike timings as well as the \textbf{Pearson correlation coefficient}. For the latter, we subsample 200 randomly chosen neurons in each population and analyse spike trains binned at \SI{2}{\milli\second} intervals.

\section{Load Balancing}
\label{app:loadbalance}
The bulk-synchronous computation scheme of the IPU requires all cores to sync after computation, meaning cores that consistently see a higher workload will cause undue idle time. The aggregation and route step are both dependent on the number of spikes per source group and therefore stand to benefit from neuron placement that takes the group-based routing scheme into account.
Outgoing connection density and firing rate are strongly correlated with neuron populations, which is why the naive round-robin placement of neurons often achieves suboptimal load balancing when a small ($N_{\mathrm{pop}}\ll C$ ) but highly active population is primarily mapped into one or very few source groups. Furthermore, because each core in our delivery scheme holds the synapse data for a unique $(\mathcal{G}_{\mathrm{src}, i}, \mathcal{G}_{\mathrm{tgt}, j})$ group combination, highly heterogeneous models in terms of outgoing connectivity may run out of memory on select cores using global round-robin.

\begin{figure}
    \centering
 \includegraphics[width=1.0\textwidth]{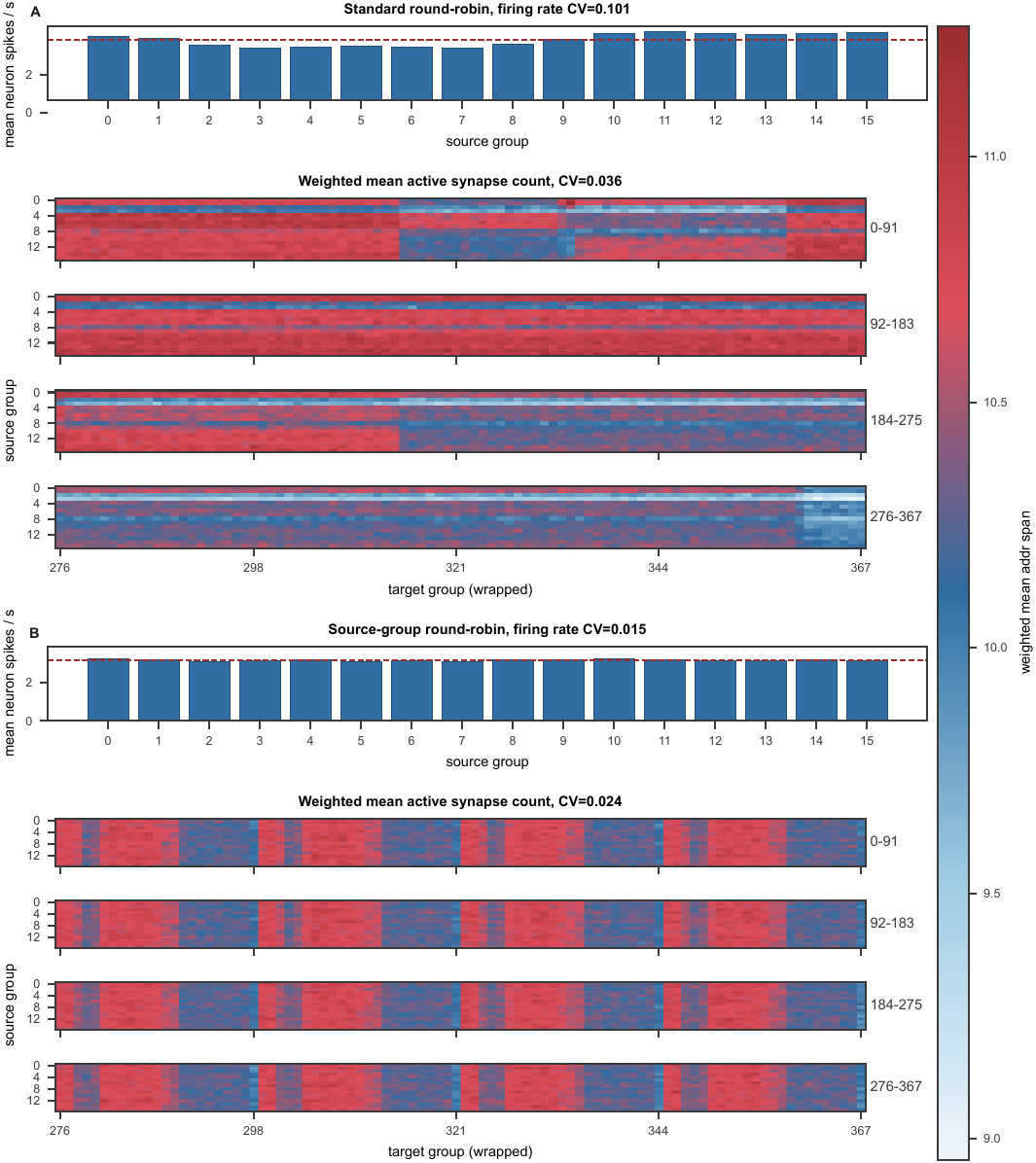}
\caption{
Distribution of mean per-neuron firing rate across source groups and the mean active synapse count (the expected number of targets) for a source neuron on any $(\mathcal{G}_{\mathrm{src}, i}, \mathcal{G}_{\mathrm{tgt}, j})$ combination. This is shown in \textbf{A.} for round-robin across all cores and \textbf{B.} round-robin within source group boundaries, respectively. All data measured over one second of biological time on identical simulation seeds.}\label{fig:loadbalance}
\end{figure}

A population- and group-aware approach therefore yields fewer stalls, a more even memory distribution across cores and, in some cases, allows for smaller buffer sizes in aggregation due to a lower maximum spike count per source group in each time step. We apply a simple source group round-robin scheme, in which neurons are placed interleaved among source groups rather than filling up cores one after another (Fig.~\ref{fig:loadbalance}). This massively improves source group spike rate balance and reduces the spread in the number of synapses accessed across cores during routing. This enables smaller static buffer sizes, which is the primary source of speedup from the source group round-robin scheme. While the balanced placement would suggest a more noticeable acceleration factor, actual performance objectives show that spike count extrema are preserved in both versions and merely more evenly distributed in the source group balanced case. This most likely comes from the populations of the cortical microcircuit model already being close in mean firing rate, along with single-neuron-level bursts and differences in outgoing connectivity that were not balanced for in this scheme.  

\section{Supporting Performance Data}
\label{app:extended_comptable}
Tab.~\ref{tab:montecarlo} shows the predicted values for active synapses and spike counts across group configurations as used in Fig.~\ref{fig:performance}e--f. Tab.~\ref{tab:competition_full} lists additional RTF runs, including process node design as well as the external drive provided. Replacing the Poisson background current with an equivalent DC compensation preserves output statistics in the PD14 model~\cite{potjans2014}.

\input{material/mc_table}
\input{material/comp_table}

\end{appendices}

\end{document}

%% file: material/mc_table.tex
\begin{table}[t]
\centering
\caption{Monte Carlo estimates of spike and synaptic workload per simulation step for the grouped placement and an ungrouped reference ($G_\mathrm{src}=1$). Mean and maximum are averaged over 100 iterations of \num{1000} time steps each. For grouped configurations, spike counts are given per source group, which every thread of the matching target-group cores scans. Active synapses are counted per thread assuming a uniform split.}
\label{tab:montecarlo}
\begin{tabular}{cccccccc}
    \toprule
    & & \multicolumn{3}{c}{\textbf{Spikes}} & \multicolumn{3}{c}{\textbf{Active synapses}} \\
    \cmidrule(lr){3-5}\cmidrule(lr){6-8}
    $G_\mathrm{src}$ & $G_\mathrm{tgt}$ & mean & max & max/mean & mean & max & max/mean \\
    \midrule
    1$^{*}$ & -- & 2.57 & 10.40 & 4.05 & 2.71 & 11.80 & 4.35 \\
    \midrule
    4  & 1472 & 6.18 & 8.79 & 1.42 & 2.71 & 13.46 & 4.96 \\
    8  & 736  & 3.09 & 5.77 & 1.87 & 2.72 & 15.29 & 5.63 \\
    16 & 368  & 1.54 & 4.05 & 2.62 & 2.71 & 18.31 & 6.75 \\
    32 & 184  & 0.77 & 3.03 & 3.92 & 2.71 & 23.31 & 8.60 \\
    64 & 92   & 0.39 & 2.39 & 6.20 & 2.71 & 31.62 & 11.66 \\
    92 & 64   & 0.27 & 2.17 & 8.09 & 2.71 & 38.11 & 14.06 \\
    \bottomrule
\end{tabular}
\par\smallskip
{\footnotesize $^{*}$ Hypothetical ungrouped synapses baseline, drawn uniformly from all $T$ threads.}
\end{table}

%% file: material/comp_table.tex
\definecolor{cpu}{HTML}{F4C2C7}
\definecolor{gpu}{HTML}{B8DDE8}
\definecolor{spinnaker}{HTML}{C8E0E8}
\definecolor{fpga}{HTML}{E8F4F7}
\definecolor{ipu}{HTML}{7CBDD4}

\clearpage
\begin{sidewaystableorg}[p]
\centering
\tablebodyfont
\caption{Extended version of published benchmark results, adapted from Senk et al.~\cite{Senk_2026}. Colours highlight similar hardware: CPU (pink), GPU (blue), FPGA (light blue), SpiNNaker (pastel blue) and IPU (dark blue).}
\label{tab:competition_full}
\begin{NiceTabular}{cccccccc}[hvlines]
    \CodeBefore
        \rowcolor{cpu}{2,9,10}          
        \rowcolor{spinnaker}{4,5}       
        \rowcolor{gpu}{3,6,7,11,12}     
        \rowcolor{fpga}{8,13,14}        
        \rowcolor{ipu}{15}              
    \Body

    \toprule
    \textbf{RTF} & \textbf{Simulator} & \textbf{Ref.} & \textbf{$E_{\mathrm{syn}}$ (\si{\micro\joule})} & \textbf{\# Nodes} & \textbf{System} & \textbf{Process (nm)} & \textbf{Drive} \\
    \midrule
    2.46  & NEST CPU      & \cite{spinnakerpd14}           & 9.94  & 12  & 2 Intel Xeon E52680v3 & 22  & DC      \\
    1.83  & GeNN          & \cite{knight2018}          & 0.47  & 1   & Tesla V100            & 12  & Poisson \\
    20.00 & SpiNNaker 1   & \cite{spinnakerpd14}           & 4.4   & 6   & 48x18xARM-968         & 130 & DC      \\
    1.00  & SpiNNaker 1   & \cite{spinnakerRealtime}       & 0.601 & 12  & 48x18xARM-968         & 130 & DC      \\
    1.05  & NEST GPU      & \cite{neuronGPU}               & 0.25  & 1   & RTX 2080 Ti           & 12  & Poisson \\
    0.70  & GeNN          & \cite{knight2021pygenn}        & {-}   & 1   & Titan RTX             & 12  & Poisson \\
    0.25  & CsNN          & \cite{heittmann2022simulating} & 0.284 & 345 & IBM INC-3000          & 28  & Poisson \\
    0.53  & NEST CPU      & \cite{Kurth_2022}              & 0.48  & 2   & 2 AMD EPYC Rome 7702  & 7   & DC      \\
    0.67  & NEST CPU      & \cite{Kurth_2022}              & 0.33  & 1   & 2 AMD EPYC Rome 7702  & 7   & DC      \\
    0.38  & NEST GPU      & \cite{gpuRuntimeConstr}        & 0.104 & 1   & RTX 4090              & 5   & DC      \\
    0.27  & GeNN          & \cite{gpuRuntimeConstr}        & 0.074 & 1   & RTX 4090              & 5   & Poisson \\
    0.05  & neuroAIx      & \cite{kauth2023neuroaix}       & 0.048 & 35  & NetFPGA SUME          & 28  & DC      \\
    0.79  & horiz/mono/s  & \cite{lindqvist2024fast}       & 0.021 / 0.009 & 1 & 1 Agilex 7 FPGA & 10 & Poisson \\
    \textbf{0.044} & \textbf{IPU} & \textbf{This work} & \textbf{0.013} & \textbf{4} & \textbf{Graphcore Bow IPU} & \textbf{7} & \textbf{DC} \\
    \bottomrule
\end{NiceTabular}
\end{sidewaystableorg}